\documentclass[usegraphicx,useAMS,usenatbib]{mn2e}

\title{The non-ballistic  superluminal motion in the plane of the sky-II}

\author[Biping Gong]
    {Biping Gong$^{1}$\thanks{E-mail: bpgong@mail.hust.edu.cn} ,  S. W. Kong$^{2}$, F.
Xue$^{1}$, Yaping Li$^{1}$, Y. F.Huang$^{2}$\\
$^1$Department of Physics, Huazhong University of Science and
Technology, Wuhan 430074, China,  \\
$^2$Department of
Astronomy, Nanjing University, Nanjing 210093, China \\
}

\begin{document}

\date{Accepted, Received ;}

\pagerange{\pageref{firstpage}--\pageref{lastpage}} \pubyear{2002}

\maketitle

\label{firstpage}

%\title{The ballistic and non-ballistic jet motion in  AGNs and X-ray
%binaries}
%%%%%%%%%%%%%%%%%%%%%%%%%%%%%%%

%$^2$ Particle Astrophysics Center, Institute of High Energy Physics,
%Chinese Academy of Sciences, Beijing 100049, China \\
%$^3$ Urumqi Observatory, National Astronomical Observatories, CAS,
%40-5 South Beijing Road, Urumqi 830011, China \\}

\begin{abstract}
{The model of non-ballistic jet motion proposed in 2008 provides a simple
explanation to the inward jet motion and bent jet.  Recently, evidences of such a non-radial motion increase rapidly, and more complicated morphologies appear.
On the other hand, the ballistic plus precession model likely holds in majority samples of jet motion.
This paper discusses the relationship between the ballistic and non-ballistic model of jet motion, which  suggests that the interaction of  ejectors with ambient
matter can produce knots at different stages of evolution and hence
different separations to the core. And as a jet precesses,
knots produced between the core and the deceleration radius result
in spiral pattern expected by the model of ballistic plus precession; and knots
generated at the deceleration radius display non-radial motion such
as bent jet or oscillation of ridge-line.
This paper develops the first non-ballistic model in four aspects. Firstly, it provides a numerical simulation to the production of multi-knot for a
precessing jet. Secondly, it fits the precession behavior of
multi-knot and interprets the oscillation of ridge lines like S5 1803+784. Thirdly, it gives an unified interpretation to the bent jet applicable to both  multi-knot and single knot. And fourthly, the problem of very large numbers of
observed outward motions as opposed to the inward ones is addressed in a new scope.
}
\end{abstract}


\section{Introduction}

To explain phenomena of non-radial jet motion, which are difficult to understand under the context of  ballistic motion of out flow~\citep{Rees66},
the non-ballistic model was proposed~\citep{Gong08}, in which a continuous jet
produces a discrete hot spot by the interaction of jet-matter at the deceleration radius. And the precession of such a hot spot
in the plane of the sky interprets the inward and bent motion of
AGNs~\citep{Kellermann04,Agudo07}.

Recently, more and more outflows showing non-radial motion
were observed. For example, the MOJAVE program, which is intended to
investigate the parsec-scale jet kinematics of
a complete flux-density-limited sample of 135 active galactic nuclei
(AGN), has revealed many unusual properties of AGN jets.
Interestingly, about one third of the outflow components have velocity
vectors that do not point back to the core feature~\citep{Homan09,Lister09}.

Complicated non-radial phenomena were also observed in a
set of interesting BL Lac Objects e.g., BL Lac~\citep{STL03}, 0716+714~\citep{Britzen09},
1803+784~\citep{Britzen10a} (1803 hereafter) and
0735+178~\citep{Britzen10b}, which display oscillation behavior
of ridge lines. While it is usually expected that the outflow
components should escape from the central core at apparent superluminal speeds
of about 5 --- 30$c$, the new observations actually show that in about 20 years or  longer, the knots
appear stationary with respect to the core while exhibiting significant variation in their position angles.
As a result, the ridge lines of 1803 change substantially.
Such an oscillation of  jet ridge
line associating with the production  of multi-knot in the jet-matter
interaction  has not been addressed in the non-ballistic model~\citep{Gong08}, which needs to be developed.

Interestingly, the non-radial motion displayed in AGNs is also
observed in the famous X-ray binary system of SS 433, where
a pair of reversely moving, mildly relativistic jets
are detected at X-ray, optical and radio wavelengths~\citep{Margon84}.
In the 2001 observation campaign on the arcsec-scale X-ray jets\citep{Migliari05},
a knot (knot A) brightened from May 8 to May 10, and then
seemed to be moving along a precession trace
to a new place (knot B) on May 12. Again, the behavior of these two knots is
not easy to understand in the framework of  the standard ballistic
model. If both knot A and B are produced by a single
outflow traveling from A to B, then the position angle should not
change so dramatically as observed, and the travel time
also should be much longer than the reported 2 d. On the other
hand, if they are produced by two discrete outflows,
then B should be ejected earlier than A, and should correspondingly
appear earlier. This is contrary to the observations.
Moreover, the ballistic scenario predicts that both knot A and B should
move outward as a whole, which is also not observed.

In some cases the jet motion like knots A and B of SS433, have been used
as evidences  contradicting to the scenario of jet precession. In fact, the problem is
originated from the ballistic assumption rather than the jet
precession. Therefore, the relationship between non-radial and radial jet motion needs to be discussed.

Non-radial jet motions are difficult to understand in the context of
ballistic motion of outflows~\citep{Rees66}, but they can be reasonably
explained by the non-ballistic model as proposed by Gong (2008), who argued
that a continuous jet can produce discrete hot spots at the deceleration
radius due to its interaction with the ambient matter. The precession of
the jet can then lead to apparent motion of the hot spots
on the sky plane, which explains the inward and bent motions of some
AGN knots.

In this paper, we go further to show that the interaction between the jet and ambient
matter can give birth to a number of knots along the outflow. The knots can be
generated either at the deceleration radius or much closer to the core.
Knots produced near the deceleration radius can naturally explain
the variation of ridge lines as observed in BL Lac~\citep{STL03} and
1803~\citep{Britzen10a}. On the other hand, knots produced closer to the
core can explain the spiral pattern when the precession of the jet is
considered.
The structure of our paper is as follows. In Section 2, the difference
and relation between the shock-in-jet model (Marscher \& Gear 1985)
and the non-ballistic model is addressed. The mechanism of producing
multi-knots is described in detail. In Section 3, the model is engaged to
explain the oscillation of ridge lines of 1803 and BL Lac~\citep{STL03},
as well as the non-radial motion of SS 433 knots.
In Section 4, we discuss the common mechanism underlying
these sources. A simple formula is derived to describe the curvature
of non-radial motions, which provides a general insight into the non-radial
behavior displayed in various AGN sources~\citep{Kellermann04,Homan09,Lister09}.
Finally, Section 5 focus on the following interesting problem: why
the number of outward-moving knots are much larger than that of the inward-moving ones.

%%%%%%%%%%%%%%%%%%%%%%%%%%%%%%%%%%%%%%%%%%%%%%%%%%%%%%%%%%%%%%%%%%%%%%%%%%%%%%%%%%%%%%%%
\section{The non-ballistic model in multi-knot case}

\subsection{Shock-in-jet model and  non-ballistic model}

The flux density of a knot (knot $i$) can be
expressed as
\begin{equation}\label{L}
S_{\nu}(\nu,R_i, \theta)=d^{-2}_LD(R_i,
\theta)^{3-\alpha}j^{'}_i(\nu)V^{'}_i ,
\end{equation}
where  $\theta$ is the viewing angle between the jet axis and
our line of sight (LOS), which causes variation in the
flux density due to the Doppler boosting effect.
The Doppler factor is given by $D(R_i,
\theta)=\gamma_b^{-1}(R_i)[1-\beta_b(R_i)\cos\theta]^{-1}$,
with the Lorentz factor $\gamma_b(R_i)=[1-\beta_b^2(R_i)]^{-1/2}$.
$V^{'}_i$ in Eq. (1) is the volume in the co-moving  frame of
knot $i$, which is connected with the volume in the observer's frame by
$V_i=DV^{'}_i$. $j^{'}_i$ is the emissivity, which  can be
derived by considering the detailed mechanism.
There are mainly three kinds of models to explain the outflow radiation,
the shock-in-jet model (Marscher \& Gear 1985), the internal shock
model (Spada et al. 2001), and the non-ballistic model (Gong 2008).
We discuss them one by one below.

The shock-in-jet model~\citep{MG85} explains the light-curve
(L-C) by assuming that each peak in the L-C is associated with the
evolution of a single shock front. In this model, the shock wave
propagates along the jet, experiencing three different regimes sequentially,
i.e. the inverse Compton loss regime,  the synchrontron radiation regime, and finally
the adiabatic regime. At each stage, the emissivity, $j^{'}_i$, can be calculated when the
parameters of radius, $R_i$, bulk velocity, $\beta_b(R_i)$, matter density,
magnetic field, etc, are known.

On the other hand, the internal shock model of gamma-ray bursts (GRBs) has been
generalized to account for the outflow evolution. In this model~\citep{GRBint}, the central engine producing the outflows is assumed to work intermittently,
so that later but faster shells can catch up and collide with slower earlier ones,
producing strong shocks. The shocks can convert a part of the bulk kinetic energy
into internal energy of electrons which then can be dissipated via various
radiation mechanism to manifest as bright knots.

Note that the knots produced in both the shock-in-jet model and the internal
shock model should be very close to the core, at least much closer as compared with the
deceleration radius. However, the deceleration radius, used by the
non-ballistic model, is a natural consequence of the dynamical interaction between
the outflow and the circumambience matter. It is very similar to the external shock
model widely used in investigating the afterglows of GRBs. Assuming that
a beamed outflow with a kinetic energy $E$ is ejected from the core. It expands into
the ambient medium with a density of $n_0$. The supersonic motion of the ejecta should
drive a blast wave propagating into the interstellar medium (ISM). At the same time,
the ejecta itself should be decelerated due to more and more swept-up material.

The evolution of the outflow can be calculated by considering the
conservation of energy~\citep{wang03,huang00}, which can be simplified as,
\begin{equation}
\label{GRB-1} E_{shock_1}(t_1)+E_{shell_1}(t_1)=E(t_1) \,. \ \
\end{equation}
where $t_1$ is the starting time of the activity, $E(t_1)$ and
$E_{shock_1}(t_1)$ are the total energy of the
outflow and the kinetic energy of the shock, respectively,
and $E_{shell_1}(t_1)$ is the internal energy of the shock.
The deceleration radius is defined as the radius where the mass of the swept-up
matter equals to $1/\gamma_b$ of the initial mass of the outflow. The external shock,
and consequently the emissivity, should be the strongest at the deceleration radius.
So it also determines the positions of the observed knots.
Therefore, in the non-ballistic scenario, the
bulk velocity, $\beta_b(R_i)$, $R_i$, and $j^{'}_i$ of
Eq.~($\ref{L}$), can be obtained from the dynamics described by
Eq.~($\ref{GRB-1}$).
Note that the deceleration radius is mainly determined by the energetics
of the outflow and the density of the ambient medium. It means
that the distance between the knots and the central core should also be
relatively fixed.

The shock-in-jet model can produce a sequence of knots by adopting
repeated onsets of several shocks. Consequently, when the jet
precesses, a spiral pattern consisted of the knots may be formed.
It is very similar to the ballistic plus precession scenario previously
used to explain the morphology of SS 433. In other words, the
shock-in-jet model itself is one kind of ballistic plus precession
model, owing to the bulk speed of the knots.

In contrast, in the non-ballistic model, a knot should stay at
approximately constant separation (at the deceleration radius)
from the core~\citep{Gong08}.  When the jet precesses, a sequence
of knots can also be produced, but all the knots should be static.
They should not move significantly outward. Of course, since
the matter density at different directions may differ, the
separations of the knots from the core is not necessarily a constant
at different directions. This provides a possibility of varying the
input parameters when we calculate the flux
of a knot by using Eq.~($\ref{L}$).

In short, the non-ballistic model differs from the shock-in-jet
model in at least two aspects: (i) It can produce knots that
reside at deceleration radius. (ii) It can naturally generate
a ring-like trace, instead of a helical one as predicted by the
shock-in-jet model.

\subsection{The production of multiple knots during jet precession}

Eq.~($\ref{GRB-1}$) is applicable for the continuous jet of AGNs and
X-ray binaries thanks to the precession of the jet.
The active timescale of the central engine is usually negligible
as compared with the precession period. For example, for a precession
period of $10$ years, the typical burst timescale of
$\delta t=1$ks is sufficiently short, during which the change of
the jet precession phase is only $\delta\eta\sim 3\times
10^{-6}$rad. In such a short time scale, the energy ejection can
be regarded as instantaneous and the interaction with ambient matter
can be treated by using Eq.~($\ref{GRB-1}$). It will appear in the
sky as a bright knot.

At another moment of $t_2=t_1+\delta t $, another ejecta comes out at a
slightly different direction with respect to the LOS, $\theta_2=\theta_1+\delta\theta$.
Such a process can occur repeatedly, giving birth to multiple knots.

%%%C1

If the jet is  continuous and with exactly constant power,
$E(t_i)=$const,  and the surrounding ISM is homogeneously
distributed, then the process described by Eq.~($\ref{GRB-1}$)  may perform  numerous times (e.g.,
10yr/1ks=$3\times 10^5$), and knots should be produced farther and
farther away from the core. However, many AGNs show  variability of
time scale of $\sim 10$ks ~\citep{Gliozzi04,Rani10}. The beginning of
such a pulse, with a period of $\sim 10$ks, means restarting the process of Eq.~($\ref{GRB-1}$)
again.

Consequently, during  a pulse (not necessarily a constant), a
numbers of bursts, say, 10, are generated, and  only a few of them
can reproduce L-Cs as shown in Fig.~1. Other bursts may produce knots
at places deviating significantly from the deceleration radius or
with flux density much weaker than that in Fig.~1. Thus,
statistically the pattern formed at each pulse is stable, in which
the knot-core distance of a knot is approximately constant or
constrained in a certain range. A schematic demonstration
of such knot production is shown in  Fig.~2, in which X1-X3 correspond to knots
produced in one pulse, and Y1-Y3 by another pulse. The upper panel
differs from the bottom one in precession phase, the former has
small discrepancy and the latter one large.

Therefore, at a direction slightly differs from the previous one, once
the energy ejection of the pulse  and the matter density etc
satisfy the parameters of Table~1, knots with  L-Cs as shown in Fig.~1 can be reproduced again.
The small phase discrepancy  among the knots produced in one pulse of the central engine results in the pattern like, knots X1-X3, and large phase discrepancy corresponds to the pattern of Y1-Y3 in Fig.~2.

On longer time scale, e.g., with $10^3$ pulses,  the change of jet
precession phase is of $\sim 10$deg, during which the matter density
can be  variable through e.g., evacuated bubbles around the sources
due to previous activities of the jets. By Eq.~($\ref{GRB-1}$) such
a fluctuation in matter density makes the L-C peak  at different
distance to the core as shown in Fig.~1, which varies the knot-core
separation of  knots at different directions.
Thus, the ring through X1 and Y1, as shown in Fig.~2, can be tilted, so are rings through X2,Y2 and X3,Y3.

Therefore,  the formation of multi-knot morphology can be realized by
extending the former non-ballistic model, where  several knots are produced along
the jet axis. And by the jet precession, these knots cause ring like
traces  at different separation to the core. Observing at different
epochs result in the oscillation of ridge lines.

%%%%%%%%%%%
The ring like pattern corresponds to the knot produced at the
deceleration radius. Comparatively, spiral patterns correspond to
the knots produced between the deceleration radius and the core,
which can be interpreted by the shock-in-jet model.

%%%%%%%%%%%%%%%%%%%%%%%%%%%%%%%%%%%%%%%%%%%%%%%%%%%%%%%%%%%%%%%%%%%%%%%%%

Thus simply replacing the knot-core distance, $R_d$
of Gong(2008), describing  the precession of one knot under the
non-ballistic model by the multi-knot, $R^i$ (where $i=1,2,3...$
corresponds to different knots). The projection of $R_i$ to
$\Delta\delta$ and $\Delta\alpha$ axes gives,
\begin{eqnarray}\label{lk}
% R^i_{x} &=& R^i[\sin\lambda\sin I\cos\eta^i +\cos\lambda\cos I ] \,, \nonumber\\
% R^i_{y} &=& R^i[\sin\lambda\sin\eta^i ]\,, \nonumber\\
% R^i_{z} &=& R^i[\cos\lambda\sin I -\sin\lambda\cos I\cos\eta^i ]
%\,,
 R^i_{\alpha} &=& R^i[\sin\lambda\sin\eta^i ]\sin\xi+R^i[\cos\lambda\sin I -\sin\lambda\cos I\cos\eta^i ]\cos\xi   \,, \nonumber\\
 R^i_{\delta} &=& R^i[\sin\lambda\sin\eta^i ]\cos\xi-R^i[\cos\lambda\sin I -\sin\lambda\cos I\cos\eta^i ]\sin\xi
\,,
\end{eqnarray}
where $\lambda$ is the opening angle of the precession cone, $I$ is
the inclination angle between the jet rotation axis and LOS. The
precession phase is $\eta^i=\dot{\Omega} t+\eta^i_0$ ($\dot{\Omega}$
is  the precession velocity of the jet, $\eta_0$ is the initial
phase of each knot). Projecting $R^i$  into the coordinate system
$x-y-z$, where the $x$-axis is towards the observer. Then rotating around
the $x$-axis for angle $\xi$, so that the new $y$-axis
($\Delta\delta$) will point north, and the new $z$-axis
($\Delta\alpha$) will point east. The position angle of a knot can
be simply obtained by
\begin{equation}
\label{pa} \psi=tan^{-1}(\Delta\alpha/\Delta\delta) \,.
\end{equation}
%\overline{}

% in the two extreme case exreme rel and non re analtical
%% j can be given, but in our case of tran-rel, numerical only (ref)
%or longair?,

\section{Application to AGN and X-ray binary}

%\subsection{Quasar 1928+738}

\subsection{Oscillation of ridge line}

The blazar S5 1803+784 is a flat-spectrum radio source at high  declination~\citep{Witzel87}. Geodetic and astronomical VLBI data gathered at
8.4 and 5 GHz between 1979-1987 showed that the component located at 1.4 mas from the core appears stationary~\citep{SLS88,Witzel88}. The stationary component was found to have non-constant core separation or oscillatory type behavior~\citep{Britzen05}.  And recently the oscillation of ridge lines of S5 1803+784 is reported~\citep{Britzen10a}.

In the following we describe the  fitting of the kinematics in the
pc-scale jet of 1803 by the extended non-ballistic superluminal model of Section~2.
The evolution of core separation, flux and position angle of one
component, C1, in the time interval between 1984-1996 are
observed~\citep{Britzen10a}, as denoted by the filled squares shown
in Fig.~3. Under the non-ballistic scenario, the behavior of C1 is
simply a conical precession of one knot,  projecting  to the plane
of sky.

With the fitting parameters of Table~1, the peaks of  L-Cs in Fig.~1 can  account for the observational knots, e.g.,
C1,Ca and C4 of 1803~\citep{Britzen10a}. By Fig.~1, at a time scale
of half a year, the peaks can decline for 20$\%$-10$\%$. More rapid
declination is expected in case of larger spectral index of
electron, $p_0$, or a radiative dynamics of the shock\citep{sari98}.
And considering the effect of Doppler boosting,  the L-C of a knot can
vary more dramatically.

The flux fitting of Fig.~3 corresponds to a bulk speed of
$\beta_b(R_i)\sim 0.3-0.6$. Therefore, the core separation, position
angle, and  flux density of C1, can be fitted by Eq.~($\ref{lk}$),
Eq.~($\ref{pa}$) and Eq.~($\ref{L}$) respectively. The variation
of separation, position angle and flux of
C1~\citep{Britzen10a}  are fitted by two groups of free
parameters. Group A contains 9 global parameters, such as precession
speed of jet axis, the opening angle of precession cone and the bulk
speed as shown in the first row of Table~2. Group B includes 13 oscillation
parameters,  denoting the deviation of the distance of component C1
to the core at 13 different epochs from the averaged value as shown
in Table~2. These two groups of parameters totally 22, fit three
figures, the evolution of the core separation, flux and position
angle at 13 epochs, totally  $13\times 3=39$  observational points.
The three fitting results are reasonable well, as shown in Fig.~3.
Notice that in the fitting of position angle and flux of Fig.~3 (26
observational points) only the 9 global parameters are needed,
as shown in the first row of Table~2.

%%%%%%%%%%%%%%%
Other components, C2,C3...can be treated similarly, except
discrepancies  on the knot-core separations and phases. The
projection of these knots at different time explains the
oscillation of ridge lines observed~\citep{Britzen10a}. Variation of
knot-core distance up to 50 percent of the average one, $R_0$, is
needed in the fitting of  Fig.~3, which indicates the fluctuation of
ISM distribution and hence large discrepancy in knot-core distances
at different directions.
%%%%%%%%%%%

The time taken by a knot to precess for tangent distance of the size
of a knot, $\delta r$, is $\delta t_{pr}=\delta
\eta_r/\dot{\Omega}$, where $\delta\eta_r=\delta r/(R^i\sin
\lambda)$. The cooling time of a knot, $\delta t_{co}$,
corresponding to, i.e., the time taken for the radio peak to decline
for a certain percentage, can be inferred from the L-C, as
shown in  Fig.~3.

\subsection{Precessing jet nozzle}

Very Long Baseline Array (VLBA) in an observing mode sensitive to
linear polarization at wavelength 7 mm with a resolution of order of
0.2mas has been performed on BL Lac at 17 regular epochs from
1998.23 to 2001.28~\citep{STL03}.

The observations suggest relatively straight trajectories near the
core, increasing in curvature at large separations to the
core(greater than 1-2 mas). The observed continuation of the
trajectory of component S10 did not fit the prediction of helical
model~\citep{Denn00}.

As shown in Fig.~4,  ridge lines formed by four knots  are observed
in December 1999, April 2000 and April 2001~\citep{STL03},  which are
labeled by points with error bars, A1-A4, B1-B4 and C1-C4
respectively.

The largest discrepancy appears in the ridge line in December 1999,
when the helical model predicts a ridge line approximately along the
vertical line from (0.0, 0.0) to (0.0, -0.6), the observed ridge
line corresponds to  points with error bars, A1-A4 in Fig.~4.

%Such an  inconsistence can be understood by the bottom panel of
%Fig.~2. By the prediction of the ballistic model, when knot X2
%reaches at a larger radial separation, $\delta R$ (at the ring lying
%knots X1 and Y1), at next time, $\delta t=\delta R/\beta c$, it
%would interact with the ring at a place between X1 and Y1, owing to
%the radial motion of knot X2 at the speed, $\beta$, e.g.,
%$\beta=0.989c$~\citep{STL03}. In fact, the observed knot is at Y1
%instead of between X1 and Y1.

By the non-ballistic model, with fitting parameters of Table~3,  the observed knots, the
ridge lines A1-A4, B1-B4, and C1-C4, can be well fitted by the filled circles, squares and triangles respectively,
as shown in Fig.~4. The relatively small discrepancies in a
few points in Fig.~4 can be improved by assuming small variation in
the core-knot separation. This means that even the straight ridge
lines can be explained by knots predicted by the non-ballistic scenario. Consequently, the true ballistic jet motion should
occur in the region even closer to the core than these straight
ridge lines of Fig.~4.

%Instead, these knots are at the stage of  the phase velocity
%and the group velocity, $v_g\ll v_p$
%which can be seen as near the deceleration radius, then
%the change of such straight ridge lines can be explained as shown in Figyy.

%Therefore,  the formation of multi-knot morphology can be understood
%in the context of GRB model, where  several knots are produced along
%the jet axis. And by the jet precession, these knots cause ring like
%traces  at different separation to the core. Observing at different
%epochs results in the oscillation of ridge lines.

%The nearly straight ridge line is due to small
%discrepancy in precession phase in these knots. and scale of radius

It was suggested that initial straight component trajectories and the
subsequent bending jet is due to a transition from a
ballistic fashion to non-ballistic flow~\citep{STL03}.

Whereas, from the point of view of non-ballistic model, the
non-radial knot can be interpreted simply by extending the fitting
of Fig.~4 to knots with larger separation to the core and with
larger discrepancy in the initial  precession of each knot, which is further discussed in Section~4.

%%%%%%%%%%%%%%%%%%%%%%%%%%%%%%%%%%%%%%%%%%%%%%%%%%%%%%%%%%%%%%%%%%%%%%%%%%%%
\subsection{Two puzzling knots of SS 433}

Interestingly, the non-radial motion displayed in AGNs is also
observed in the famous X-ray binary system of SS 433.
, in which anti-parallel jets traveling
near-relativistically are detected in observations across X-ray,
optical and radio wavelengths~\citep{Margon84}. The strong and broad emission lines have been identified as redshifted/blueshifted emission from collimated jets with intrinsic velocity of $\simeq0.26c$~\citep{Abell79}. The periodic change of Doppler shift of emission lines with time is widely accepted to be
a precessing ballistic jet with a period of 164d.

Although the jet motion of SS 433 is less than the speed of light, the behavior of which is strikingly similar to the superluminal sources.
In the 2001 observation of  arcsec-scale X-ray jets
of SS 433 \citep{Migliari05}. As shown in the top panel of Fig.~5, a
knot became brighter from May 8 to May 10 (knot A) and subsequently
on May 12 it appeared to be moving along a precession trace
to a new place (knot B).

%This indicates that the two knots are lying on the precession trace,
%at different precession phases, which brighten up sequentially.

The observation  of the two knots seems simple, it is
not easy to understand in the context of  the standard ballistic
plus precession model either.  If knots A and B are results of one
ball traveling from A to B, then its position angle should not
change so dramatically, and the discrepancy in the travel time of
these two knots to the observer should be much longer than the reported
 2 d. On the other hand, if knots A and B are caused by two
ejections, then B should be ejected earlier than A. As a result, B
should appear earlier than A. This is contrary to the observation.
Moreover, the ballistic scenario  predicts that both knots A and B should
move outwardly as a whole. Whereas, one cannot find such a motion
either.

To explain the motion of these two knots under the standard
ballistic model, the underlying faster outflow  scenario is
proposed, in which the two knots ejected from the binary core first,
and later a fast shock wave propagates from the core through the jet
and hits first the knot A and then knot B \citep{Migliari05}. This
requires the central engine to switch  on different modes of power
at different time.

%Interestingly,  the trajectory of these two knots as shown in
%Fig.~1, is strikingly similar to the non-radial trace of BL Lac
%2200+420 as shown in Fig.~5a. Again, the acceleration perpendicular
%to the velocity vector is in action.

%Apparently, it is not easily feasible to explain these non-radial phenomena within
%the scenario of ballistic motion plus precession.

In some cases the jet motion like knots A and B of SS433, have been used
as evidences  contradicting to the scenario of jet precession. In fact, the problem is
originated from the ballistic assumption rather than the jet
precession. Therefore, the relationship between non-radial and radial jet motion needs to be discussed.

The model of two knots moving at certain position and
then brightened up by two shocks  sequently~\citep{Migliari05}, is
shown in the middle panel of Fig.~5.  Knot B is at
a distance of $1.7\times 10^{17}$cm from the core, which was ejected
from the binary system about 40d earlier than knot A, and is moving
towards us with an angle to LOS of $\sim 80$deg. The shock wave
interact first with knot A and then with knot B.  When the shock
interacts with knot A, the distance between A and B is about 20
light days. Therefore, the shock wave has to travel the projected
distance between the two knots in about 22d. In order to observe the
brightening of the two components within two days, the shock has to
travel with a velocity of $v\sim 0.5$c.

The non-ballistic model provides a simpler scenario, in which knot
 A with a X-ray peak count of 75.2  at $\sim
10^{17}$cm~\citep{Migliari05} can be reproduced at the deceleration
radius $r_{\rm d}$ (measured from the central), where the swept-up
medium by the jet  has an energy comparable to that of the
outflow~\citep{bm76},
\begin{equation}
r_{\rm d}=(\frac{3 E}{4 \pi \gamma_0^2 n m_p c^2})^{\frac{1}{3}},
\end{equation}
where $\gamma_0$ is the initial Lorentz factor of the outflow. So
the separation  of knot A to the core of $\sim 10^{17}$cm corresponds to
$E_{51}^{\frac{1}{3}} \gamma_{0,5}^{-\frac{2}{3}} n_0^{\frac{1}{3}}
\sim 0.54$, where $E_{51}$ is $E$ in units of $10^{51}$ erg,
$\gamma_{0,5}$ is $\gamma_0$ in units of 5, $n_0$ is in units of 1
${\rm cm}^{-3}$.

%Moreover the radio knots with a core separation of $\sim
%10^{15}$cm~\citep{Migliari05} are likely produced by the low energy
%tail of the peaks of high energy emission.

By the non-ballistic model, knots A and B observed in  SS 433 can be fitted by the
precession trace as shown in the bottom panel of Fig.~5.  The
fitting parameters are given in Table~2. Therefore, knots A and B
can be lying on the precession trace given by Eq.~($\ref{lk}$). The
deviation in precession phase of A and B corresponding to a time
discrepancy of 38 days. It takes 36 days for a signal to propagate
from A to B. Hence knot B is brightened up two days later than knot A,
which satisfies the observation~\citep{Migliari05} in a much simpler
way.

%%%%%

The fitting of the two dimension trajectory of knot A and B by
Eq.~($\ref{lk}$) actually needs 4 equations (corresponding to the
coordinate of A and B), and the 2d deviation in propagation time
imposes another constraint on the discrepancy of the knot-observer distance.
Therefore, the bottom panel of Fig.~5 can be recognized as obtaining the five
variables, as shown in Table~4 by five equations (constraints).

Note that in principle, no matter how many input parameters are
used, the trajectory of  knot A and B cannot be fitted by the
ballistic plus precession scenario directly.

%%%%%

\section{The nature of bent jet}
The origins of  jet curvature in the one-knot case has been discussed in Gong (2008).
Here it is extended to the multi-knot case, and the association of  the two types of bent jets is addressed.

Within the non-ballistic  model, there are two origins of  jet
curvature. The first one is caused by the knots moving at different
rings. In this case, the phase discrepancies between two neighboring
knots at different separations are given, $ \delta \eta^i=
\dot{\Omega}\delta t^i +\eta^0$, in which $\delta t^i$ can be
written,
\begin{equation}
\label{phase}  \delta t^i\equiv t^{i+1}-t^i= \delta R^i/
v^{i+1}+[x^{i+1}-x^i]/c \,,
\end{equation}
where  $x^i$ is the knot-observer distance of knot $i$, $\delta
R^i\equiv R^{i+1}-R^i$ is discrepancy of  the core-knot distance
between knots $i$ and $i+1$, and $v^{i+1}$ is the bulk speed of knot
$i+1$. A dramatic reduction of bulk velocity between e.g., knot 2
and 3, $v^{3}<<v^2$, results in $\delta R^2/ v^{3}\gg \delta R^2/
v^{2}$, which leads to a large phase  discrepancy, $ \delta
\eta^2$ between knot 2 and 3. In such case, the jet is bent sharply
at knot 2 by Eq.~($\ref{phase}$) and Eq.~($\ref{lk}$).
This explains the wiggling structure
of 0548+165~\citep{mantovani98}, and the two rectangular pattern in
BL Lac object PKS 0735+178~\citep{Britzen10b}.

%Moreover, with the
%precession of jet, the phase discrepancy varies with time, so that
%rectangular pattern can appear and disappear at different
%epoches~\citep{Britzen10b}.

Knots produced in the case of  small phase discrepancy and large
phase discrepancy can be demonstrated by the upper and bottom
pattern of Fig.~2 respectively, which explains the variation of ridge lines of   BL Lac~\citep{STL03} and
the oscillation of ridge lines of 0735+178 and 1803 respectively.

%Hence, the multi-knot behavior of
%0735+178 is actually another exhibition of  the oscillation of ridge
%lines of 1803~\citep{Britzen10a} and BL Lac~\citep{STL03}, except for different precession parameters as shown in
%Table~1.

The second origin of  jet curvature corresponds to  the trajectory
produced by a singular knot, with  negligible variation in radial
separation, which is simply part of an ellipse in
the plane of the sky, which has been discussed~\citep{Gong08}.

From the stand of  Eq.~($\ref{phase}$), the  curvature in such  one-knot
case can be obtained simply by removing  the first term at the
right hand side of Eq.~($\ref{phase}$),  which well
explains the curvature of knots A and B in SS 433.

In fact, such a singular knot curvature determined by
Eq.~($\ref{phase}$) and Eq.~($\ref{lk}$) can explain the non-radial
behavior  revealed in other AGN sources,  where outflow are aligned
with the local jet direction, suggesting the jet flow occurs along
preexisting bent channels, like 0738+313 and quasar NRAO
150~\citep{Kellermann04,Lister09,Agudo07}.

By the time  derivatives of Eq.~($\ref{lk}$), $\ddot{R}^i_{\alpha}$
and $\ddot{R}^i_{\delta}$ can be obtained, which naturally provides
the acceleration perpendicular to the velocity vector displayed in
AGN components~\citep{Lister09}.

%Moreover the interpretation of the
%correlation of core-knot separation with flux of 1803, as shown in
%Fig.~6, provides a clue to  the statistics of the maximum apparent
%jet speed versus VLBA luminosity~\citep{Lister09}, which  likely
%corresponds to the time derivative of Eq.~($\ref{lk}$) versus
%Eq.~($\ref{L}$).

%%%%%%%%%%%%%%%%%%%%%%%%%%%%%%%%%%%%%%%%%%%%%%%%%%%%%%%%%%%%%%%%%%%%%%%
%%%%%%%%%%%%%%%%%%%%%%%%%%%%%%%%%%%%%%%%%%%%%%%%%%%%%%%%%%%%%%%%%%%%%%%

\section{Why 98.5$\%$ outward motion}

Although  the problem that the
jet features showing an overwhelming tendency  of outward
motion  away from the core has been addressed~\citep{Gong08}, it can be further discussed in  three aspects.

%In such cases, one cannot see a knot
%moves inwardly to the core, owing to these knots are far away from
%the core. In other words,
%Such wiggling motion is usually not
%cataloged to the inward motion, although  the precession mechanism
%is  the same.

%The third is that the geometry of precession results in the wiggling
%pattern as quasar 1928+738~\citep{Hummel92}, in which the outward
%motion of jet corresponds to small misalignment angle with the LOS,
%while the inward jet motion has larger angle and hence Doppler
%deboosted significantly.

The first is that most jets of AGNs and X-binaries are observed at
the stage before ejectors reaching the deceleration radius, which is
dominated by the ballistic scenario.

The second is that some  sources  have a small opening angle of
precession cone, the geometry of which displays  the oscillation of
ridge line like BL Lac~\citep{STL03}. In such case, outward and
inward motions are performed in  limited region, and motion
perpendicular to the ridge line is obvious, which is usually
cataloged to the helical motion instead of  inward motion.

The third one is for those sources that the jet axis really through
LOS during the precession of jet. In such case,  following mechanism
can make asymmetry in the inward and outward motion of such a precession
jet.

By Eq.~($\ref{phase}$),  a mildly curved jet appears when $\delta
R^i/ v^{i+1}$ is not too large at different locations of the jet. As
shown in Fig.~6, a knot is produced at the deceleration radius at a precessing and curved jet. Since the knot actually
corresponds to a region, e.g., the knot C1 of 1803 is about $\delta
r\approx 0.5$ mas(3pc) in diameter~\citep{Britzen10a}. Owing to the
energy dissipation at the deceleration radius, the bulk speed of
ejector reduces to typically half of its original speed,
which can be simplified as composed of several sub-knots, each
corresponds to a sub-jet as $a$ and $b$  in the left panel of
Fig.~6. According to the standard GRB model, a  dramatic reduction
of bulk speed occurs at the deceleration where the knot produced
(e.g., from $\gamma=10$ to $\gamma=2$), so that the upstream
sub-knots posses larger bulk speed than those of the downstream ones,
$\gamma_i>\gamma_{i+1}$, by a factor of a few. Hence through
$\theta_i=1/\gamma_i$, the half opening angle of emission beam of
these two sub-knots, $i$ and $i+1$, satisfy,
$\theta_i<\theta_{i+1}$. Thus the effective emission beam of the
knot is extended, with higher brightness at upstream region and
lower at the downstream region, as shown in the middle of Fig.~6.

With $\alpha$ represents the misalignment angle between two sub-jets
$a$ and $b$ corresponding to the two knots having the minimum and
maximum   beam opening angles, $\theta_{min}$ and $\theta_{max}$.
E.g., for knot C1 of 1803, $\alpha\approx \delta
r/R_0=0.5/0.8=0.6$rad. As shown in the left panel of Fig.~6, a knot
beam extended significantly when
$\alpha\gg\theta_{min}$(e.g.,$\theta_{min}\approx 0.1$, with
$\gamma\approx 10$); and a negligible extension occurs when
$\alpha\approx\theta_{min}$ (e.g.,$\theta_{min}\approx 0.5$, with
$\gamma\approx 2$).

The precession of the curved jet of the left panel of Fig.~6, is
equivalent to a fix jet and a change of the observer's LOS, as shown
in the middle of Fig.~6.  At epoch 1, the observer sees a brief
inward motion, and epochs  2-4 correspond to the long outward motion
due to the extended emission beam, as shown in right panel of
Fig.~6. Thus the very large numbers of  observed outward motions as
opposed the inward ones results when  most of the knot emission
beams are deformed considerably. Consequently this causes serious
asymmetry at two sides of the line connecting the observer, the knot
and the core (case 2 in the middle and right hand side of Fig.~6).

Notice that according to this scenario, the size of an inwardly
moving knot is not necessarily large than that of an outward one. Whereas,
the explanation of inward motion by jet bending back and across
LOS~\citep{Marscher91} predicts larger size of inward knot than that
of outward ones, considering the lateral expansion of a knot at very
longer distance from the core.

In this paper, multi-knot corresponds to a series of knots produced by
jet-matter interaction.  Singular knot is an individual knot
produced by the jet-matter interaction. And sub-knot represents a
singular knot (or one of the multiple knots) at different stage of evolution near the deceleration radius.

\section{Discussion and conclusion}

This paper shows that generally the interaction of jet and ambient
matter can reproduce a number of knots along the jet, both at the
region near the deceleration radius and at the region between the
deceleration radius and the core.
 Knots produced at the region near the
deceleration radius are simulated by the dynamics of jet-matter interaction.
And the motion of such multi-knot is used to explain
the changes of ridge line such as  BL
Lac~\citep{STL03} and 1803~\citep{Britzen10a}. On the other hand, knots produced between the deceleration
radius  and the core can be understood in the context of
the Shock-in-jet model~\citep{MG85}, in which a spiral pattern is expected when the jet precesses.
In other words, this paper not only extends the former non-ballistic model from  one-knot
to multi-knot, but also discusses the relationship between the non-ballistic model and the Shock-in-jet model.

Interestingly,  jets  observed by the early long base line technique
e.g., in 1970s-1980s, appear more straight than those observed in the
past ten years with much improved sensitivity. From the view of
non-ballistic model, this is expected. Because even a curved jet can
behave as a straight one, due to the low sensitivity observation
favors to observe bright knots having small misalignment between the
jet axis and LOS, which is Doppler boosted strongly. In other words,
it is this  selection effect that makes the early jets appear
straight. By the improved sensitivity, weaker and weaker part of jet
is measurable, so that more and more curved jet appear.  Thus, the
ratio of non-radial over radial motion of jet should increase with
time.

%\clearpage

%%\begin{thebibliography}{}

\section{Acknowledgments}
We thank S. Britzen, A. Witzel, and J.A. Zensus for  helpful
discussions and suggestions in the manuscript. We also thank Z.Q. Shen for useful comments of the manuscript. This research is supported by the
National Natural Science Foundation of China, under grand
NSFC11178011. And 
11033002,  by the National Basic Research Program of China (973
Program, Grant No. 2009CB824800).

%%%%%%%%%%%%%%
\begin{table}%\scriptsize
%\tablecolumns{13} \tablewidth{0pc}
\begin{center}
\caption{The fitting parameters of three knots by the standard
afterglow of GRB model.}
\begin{tabular}{cccccccc}
\hline \hline knot & $\Gamma_0$ & $\varepsilon_e$ & $\varepsilon_B$
& $\theta_j$
  &  $E_0$ &  $n_0$ &  $p_0$  \\ \hline

1 & 5.0 & 0.1 & 0.01 & 0.01 & $6.0\times 10^{59}$ & 0.2  & 2.1 \\

2 & 5.0 & 0.1 & 0.01 & 0.02 & $2.0\times 10^{59}$ & 0.02  & 2.3 \\

3 & 5.0 & 0.1 & 0.01 & 0.03 & $8.0\times 10^{58}$ & 0.002  & 2.8 \\

\hline \hline
\end{tabular}
\end{center}
{\small The initial isotropic equivalent kinetic energy of the
outflow,
%kinetic energy of outflow,
$E_0$ (erg), opening angle of jet, $\theta_j$ (rad), initial Lorentz
factor,$\Gamma_0$, density of ambient matter,$n$, spectral index of
electron, $p_0$, the fraction of total shock energy acquired by the
shocked electron and magnetic field, $\varepsilon_e$ and
$\varepsilon_B$ respectively. }
\end{table}

\begin{table}%\scriptsize
%\tablecolumns{13} \tablewidth{0pc}
\begin{center}
\caption{The fitting parameters of the knot-core separation,
position angle and flux density of component C1 of 1803 as shown in
Fig.~6.}
\begin{tabular}{ccccccccc}
\hline \hline $\xi$ & $\dot{\Omega}$ & $I$  & $\lambda$
  &  $\eta_0$ &  $R_0$ &  $\beta_b$ & $n$ & $S_0$\\

%  & $\eta_{00}$ & $\eta_{01}$ &  $\eta_{02}$  \\
%1.43 & 40.05 & 1.28 & 0.36 & 4.24 & 0.82  & 0.31 & 2.00 \\ \hline
1.43 & 36.3 & 1.27 & 0.28 & 4.24 & 0.82  & 0.34 & 2.00 & 0.10 \\
\hline

 $\delta R_1$ &  $\delta R_2$ & $\delta R_3$ & $\delta R_4$ & $\delta R_5$ & $\delta R_6$ & $\delta R_7$ & $\delta
 R_8$ & $\delta R_9$
 \\

-0.37 & -0.45 & 0.12 & 0.17  & -0.018  & -0.16 & -0.31 & -0.20 &  -0.05 \\

 $\delta R_{10}$ & $\delta R_{11}$ & $\delta R_{12}$ &
$\delta R_{13}$ &  &  &  \\

 0.00 & -0.046 & 0.060 & -0.37 &   &   & \\

\hline \hline
\end{tabular}
\end{center}
{\small  The average  knot-core separation, $R_0$, and  the  deviation of the knot-core distance from the average value at different time, $\delta R_i$ ($i=1-13$),  are
in unite of mas;  $\dot{\Omega}$ is in deg/yr,  $\xi$, $\lambda$, $I$ and
$\eta_0$ are in rad. The two parameters, $n$ and $S_0$ denote the index $3-\alpha$ and the flux density at the righthand side of Eq.~($\ref{L}$) respectively. }
\end{table}

%\begin{table}%\scriptsize
%\tablecolumns{13} \tablewidth{0pc}
%\begin{center}
%\caption{\bf The fitting parameters of BL Lac 2200+420. }
%\begin{tabular}{ccccc}
%\hline \hline  $\xi$ & $\eta $ & $I$  & $\lambda$
%  &  $R$ \\

% 49.9 & 332.3  & 33.8 & 27.5 & 6.1 \\

%\hline \hline
%\end{tabular}
%\end{center}
%{\small  The core-knot separation ($R$) is in mas, and all other
%parameters are in degree. }
%\end{table}

\begin{table}%\scriptsize
%\tablecolumns{13} \tablewidth{0pc}
\begin{center}
\caption{\bf The fitting parameters of BL Lac (Stirling et al. 2003). }
\begin{tabular}{cccccc}
\hline \hline  $\xi$ & $\eta^{\ast}$ & $I$  & $\lambda$ &  $R$ & $\Omega$
  \\
73.1 & $30.0^{\ast}$  & 11.6 & 2.0 & 0.54 & 0.67\\
\hline \hline
\end{tabular}
\end{center}
{\small  The core-knot separation ($R$) is in mas, the precession velocity, $\Omega$ is in deg/d, and all other
parameters are in degree. $^{\ast}$ 30.0 deg corresponds to the
initial phase of the smallest ellipse in Fig~5, other initial phases
from from to large are, 0.0deg, 20.0deg and 15.0 deg respectively. }
\end{table}

\begin{table}%\scriptsize
%\tablecolumns{13} \tablewidth{0pc}
\begin{center}
\caption{\bf The parameters derived from the fitting of the trace of
knots A and B. }
\begin{tabular}{ccccc}
\hline \hline $\xi$ & $\eta $ & $I$  & $\lambda$
  &  $R$ \\
1.0 & 60.0 & 72.5 & 30.7 & 2.0  \\
\hline \hline
\end{tabular}
\end{center}
{\small  The core-knot separation ($R$) is in arcsec, and all other
parameters are in degree. }
\end{table}

%\clearpage

%\citep{Migliari05}

%\clearpage

\begin{figure}
%\epsscale{0.6} \plotone{fig2.eps}
%\begin{figure}
%\centerline{
%            \epsfig{file=fig3.eps, scale=0.4}
%            }
%\includegraphics
%[width=0.6\textwidth]
%[7,7][500,490]
%[scale=0.8]
%[7,7][300,270] {fig2.eps}
\includegraphics[width=0.45\textwidth]{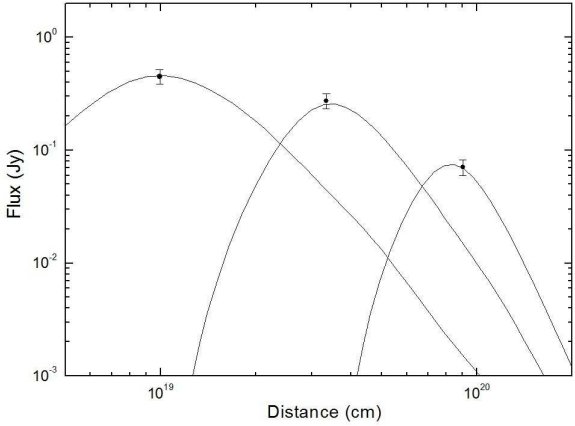}
\caption{\small The L-C of three ejecta corresponding to
Eq.~($\ref{GRB-1}$). The three peaks
corresponds to three knots at different distances to the core, C1,Ca
and C4.\label{fig1} }
\end{figure}

%\clearpage

\begin{figure}
%\epsscale{0.9} \plotone{fig3.eps}
%\begin{figure}
%\centerline{
%            \epsfig{file=fig4.eps, scale=0.4}
%            }
%\includegraphics
%[width=0.6\textwidth]
%[7,7][500,490]
%[scale=0.8]
%[7,7][300,270] {fig3.eps}
\includegraphics[width=0.45\textwidth]{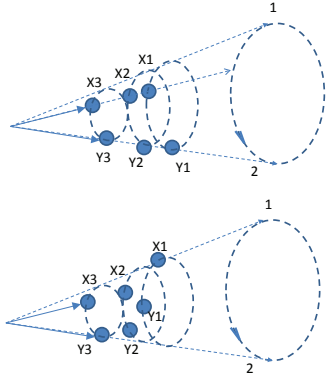}
\caption{\small A schematic figure demonstrates  knots produced by
activities  of the cental engine.  Knots X1-X3 are
produced by one pulse, and Y1-Y3 another. The upper panel corresponds to knots with small discrepancy in precession phases; and the knots of the  bottom panel differ significantly in precession  phases.
 \label{fig2} }
\end{figure}

\begin{figure}
%\epsscale{0.6} \plotone{fig4.eps}
%\begin{figure}
%\centerline{
%            \epsfig{file=fig2.eps, scale=0.4}
%            }
%\includegraphics
%[width=0.6\textwidth]
%[7,7][500,490]
%[scale=0.8]
%[7,7][300,270] {fig4.eps}
\includegraphics[width=0.45\textwidth]{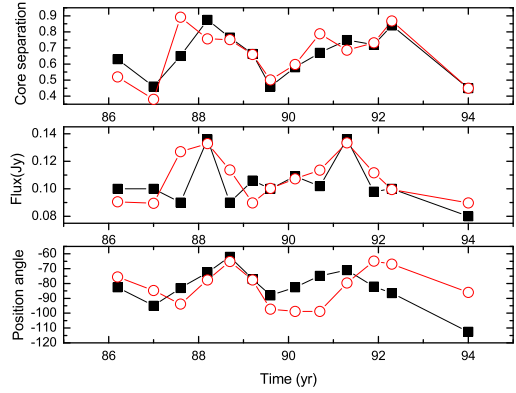}
\caption{\small The observed (filled squares corresponding to Britzen et al. 2010) and
fitted (opened circles) evolution of core separation (mas), flux and
position angle (deg) of one component, C1 of 1803, in the time interval
between 1984-1996. The corresponding $\chi^2$ from top to bottom panel are 6.3, 0.33, and 90.9 respectively.
\label{fig3}}
\end{figure}

\begin{figure}
%\epsscale{0.6} \plotone{fig5.eps}
%\begin{figure}
%\centerline{
%            \epsfig{file=fig2.eps, scale=0.4}
%            }
%\includegraphics
%[width=0.6\textwidth]
%[7,7][500,490]
%[scale=0.8]
%[7,7][300,270] {fig5.eps}
\includegraphics[width=0.45\textwidth]{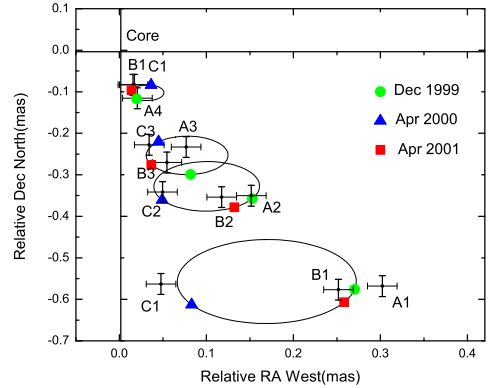}
\caption{\small  The observed and fitted ridge lines of BL Lac (Stirling et al. 2003).   Points with error bars labeled
A1-A4, B1-B4 and C1-C4 are observed in December 1999, April
2000 and April 2001, respectively. The ellipses are
predicted by the non-radial model. \label{fig4}}
\end{figure}

\begin{figure}
%\epsscale{0.7} \plotone{fig1.eps}
\includegraphics[width=0.45\textwidth]{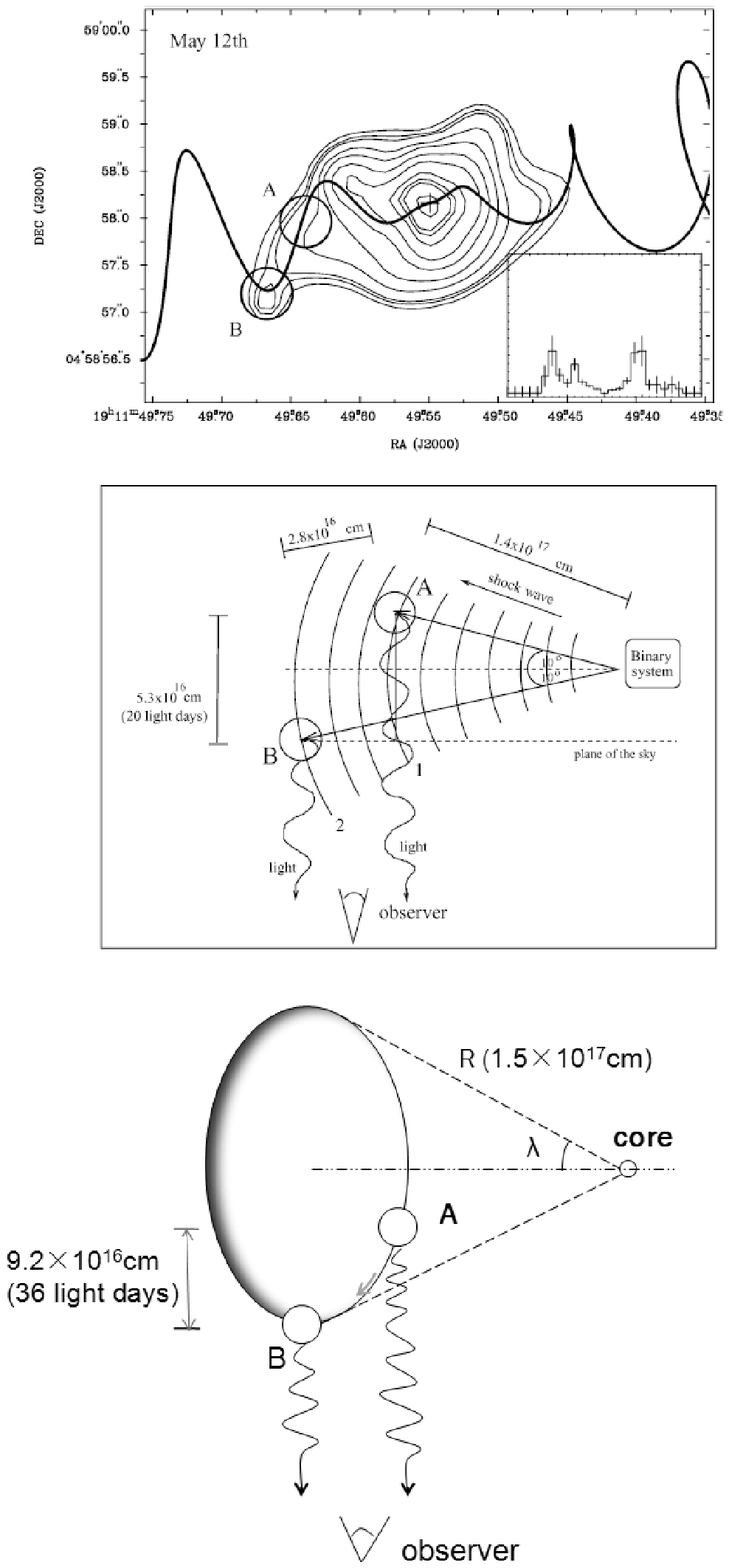}
\caption{The observation and interpretation of knots A and B of SS433.
Top panel is the zeroth-order HETGS image taken on 2001 May 12.
Middle panel is the sketch of the orientation of the east jet of SS
433, which interprets  the observations of knots A and B in context
of the slowly moving clouds energized erratically by a more
powerful, faster, unseen flow. The top and middle panels are adopted from Migliari et al. (2005).
The bottom panel
is the  sketch explaining the observations of knots A and B by the
non-ballistic model. \label{fig5}}
\end{figure}

\begin{figure}
%\epsscale{0.6} \plotone{fig6.eps}
%\begin{figure}
%\centerline{
%            \epsfig{file=fig3.eps, scale=0.4}
%            }
%\includegraphics
%[width=0.6\textwidth]
%[7,7][500,490]
%[scale=0.8]
%[3,3][720,650] {fig6.eps}
\includegraphics[width=0.45\textwidth]{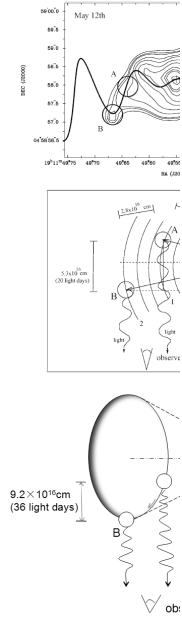}
\caption{\small A schematic explanation of the very large numbers of
observed outward motions as opposed the inward ones. The curved jet
precesses along the dash-dotted direction, and  the observed knot
can be seen as composed of several sub-knots, as shown in the left
panel. The precession of the curved jet of the left panel  is
equivalent to a fix jet and a motion of the observer's LOS as shown
in the middle panel. The right panel gives the observed knot,
hollowed circle, and the core, filled circle. \label{fig6}}
\end{figure}

\end{document}